\documentstyle[multicol,aps,prl]{revtex}
\input{psfig.sty}

   \def\A{{\cal A}}
   \def\eps{\xi}

   \def\vx{{\bf x}}
   \def\vy{{\bf y}}
   \def\vr{{\bf r}}
   
   \def\vu{{\bf u}}
   \def\vv{{\bf v}}
   \def\vde{{\bf \partial}}
   \def\de{\partial}

\begin{document}

\title{Pressure and intermittency in passive vector turbulence}

\author{L. Ts. Adzhemyan,$^{1}$ N. V. Antonov,$^{1}$
A. Mazzino,$^{2,3}$ P. Muratore-Ginanneschi$^{4}$ and A. V. Runov$^{1}$\\
\small$^{1}$
{\it Department of Theoretical Physics, St Petersburg University,
Uljanovskaja 1, St~Petersburg, Petrodvorez, 198904 Russia}\\
\small$^2$  {\it CNR-ISIAtA, c/o Lecce University, 73100, Lecce, Italy} \\
\small$^3$ {\it INFM--Department of Physics, Genova University, I--16146
Genova, Italy} \\
\small$^4$ {\it Department of Mathematics, Helsinki University,
   P.O. Box 4,  00014, Helsinki, Finland}}

\draft
\date{12 February, 2001}
\maketitle

\begin{abstract}
We investigate the scaling properties a model of passive vector turbulence
with pressure and in the presence of a large-scale anisotropy.
The leading scaling exponents of the structure functions are proven to be
anomalous.
The anisotropic exponents are organized in hierarchical families growing
without bound with the degree of anisotropy.
Nonlocality produces poles in the inertial-range dynamics corresponding to
the dimensional scaling solution.
The increase with the P\'{e}clet number of  hyperskewness and higher
odd-dimensional ratios signals the persistence of anisotropy effects
also in the inertial range.

\end{abstract}

In the last five years, much progress has been achieved in understanding
of the physical origin of intermittency and anomalous scaling in 
fluid turbulence. The study of the passive advection of a scalar quantity
(temperature, concentration of an impurity {\it etc}) by a random Gaussian
field, white in time and self-similar in space, the rapid-change model
\cite{Kraich1}, has played a crucial role in these developments.
There, it was for the first time possible to demonstrate the emergence of
anomalous scaling and to relate it to the existence of statistical
conservation laws of the dynamics \cite{Falk1,GK}. Such conservation laws
appear as scaling zero modes of the hierarchy of linear operators governing
the inertial-range dynamics of the correlation functions.
The dominance of such zero modes has been confirmed in numerical
experiments \cite{FMV} and demonstrated analytically by means of
the renormalization group and operator product expansion \cite{RG,RG1}.
The importance of these results is seen from the fact that statistical
invariants of dynamics are features of the passive advection by more general
classes of turbulent velocity fields \cite{CV}. A review and more
references can be found in Ref.~\cite{Nature}.

The first-principle model of fluid turbulence is the Navier-Stokes
equation, a nonlinear integro-differential equation for the
solenoidal vector velocity field.
Whether the relation between statistical invariants and anomalous
scaling can be generalized to the nonlinear dynamics of the
advecting velocity field remains an open and challenging issue.
Setting aside the problem of the nonlinearity of interactions,
the scope of the present Letter is to investigate how nonlocality
affects scaling.

In the Navier-Stokes equation nonlocality arises through the pressure
term fixed by the incompressibility condition for the velocity field.
Here, we consider the passive advection of an incompressible vector field
$\vu$ according to the most general dynamics consistent with the Galilean
invariance:
   \begin{eqnarray}
   \partial _t\vu + (\vv \cdot \vde) \vu - \A (\vu \cdot \vde) \vv
   + \vde {\cal P} =\kappa \Delta \vu + {\bf f} ,
   \label{equation}
   \end{eqnarray}
where ${\cal P}$ is the pressure and $\kappa$ is the diffusivity coefficient.
The advection field $\vv$ is specified by the Kraichnan ensemble. It models a
homogeneous and isotropic nonintermittent velocity field and, due to
Gaussianity, is determined completely by the second-order structure function
   \begin{eqnarray}
   S_{\alpha\,\beta}({\bf r})=\langle v_{\alpha}(\vx,t)
   \left(v_{\beta}(\vx,t')- v_{\beta}(\vy,t')\right) \rangle =
   D\ \delta(t-t')|\vr|^{\xi} [(d-1+\xi)\delta_{\alpha\,\beta}-
   \xi\,\hat{r}_\alpha\,\hat{r}_\beta]
   \label{velocity}
   \end{eqnarray}
with $\vr\,\equiv\,\vx-\vy$ and $\hat{\vr}= \vr/|\vr|$.
Here $0<\xi<2$ is a kind of H\"{o}lder exponent which measures the
``roughness'' of the velocity field. In the renormalization group approach,
it plays the same role as the parameter $\varepsilon=4-d$ in the theory of
critical phenomena.

The vector field $\vu$ is stirred by an incompressible forcing ${\bf f}$
varying only over large spatial scales. We assume the forcing ${\bf f}$
to be delta-correlated in time and Gaussian in space with the covariance
  \begin{equation}
   \langle  f_{\alpha}(\vx)\, f_{\beta}(\vy)\rangle
   = \delta(t-t')\, F_{\alpha\,\beta}\left(\vr/L\right).
   \label{force}
   \end{equation}
The parameter $L$ is the integral scale of the stirring, and
$F_{\alpha\,\beta}$ is a function which rapidly decays as
$\vr/L \to\infty$ and tends to a constant as $\vr/L \to 0$.

As in the Navier-Stokes equation, the pressure ${\cal P}$ is fixed by
requiring Eq.~(\ref{equation}) to be consistent with the incompressibility
condition:
   \begin{equation}
   \Delta {\cal P}(\vx,t) =(\A-1)\,(\partial_{\alpha} v_{\beta})
   ( \partial_{\beta} u_{\alpha}).
   \label{Poisson}
   \end{equation}
The parameter $\A$ modules the degree of nonlocality in the problem.
The case $\A=1$, known as the Kazantsev-Kraichnan kinematic dynamo model,
represents the limit of local interactions. It describes the early stage
of the growth of a magnetic seed in a turbulent fluid. However, the
exponential growth of the magnetic field occurs only if $\xi$ is large
enough, in particular for $\xi>1$ in three dimensions\cite{V96}. For smaller $\xi$,
a steady state sets in exhibiting anomalous scaling behavior already for
the pair correlation function. The corresponding exponents have been found
exactly in the isotropic \cite{V96} and anisotropic \cite{LM99,ALM99,Arad}
sectors.

As $\A$ deviates from unity, the solution of the Poisson equation
(\ref{Poisson}) introduces in the dynamics a nonlocal integral term.
Cases of special interest are $\A=0$ and $\A=-1$. For $\A=0$
the stretching term $(\vu \cdot \vde) \vv$ vanishes and the dynamics
becomes invariant under translations $\vu\to\vu+{\rm const}$.
The kinetic energy $\langle \vu^{2}(\vx,t) \rangle$ is conserved, as in the passive
scalar and stochastic Navier-Stokes equation; see Ref.~\cite{LA,Arad2}.
The case $\A=-1$ corresponds to the linearization of the Navier-Stokes
equation around the rapid-change velocity field; in two dimensions it
was studied in Ref.~\cite{YK}.

We are interested in the scaling properties of the correlation functions
of the field $\vu$ in the inertial range,
$L \,>>\,|\vr|\,>> \ell \,\equiv\,(\kappa/D)^{1/\xi}$, where $\ell$ is
the dissipative scale.
The equal-time pair correlation function is governed by a linear
{\em integro}-differential equation,
\begin{eqnarray}
\de_t C_{\alpha\,\beta}= {\cal D}_{\alpha\,\beta}^{\gamma\,\delta}
\,C_{\gamma\,\delta},
\quad
C_{\alpha\,\beta}({\bf r},t)=
\left\langle u_{\alpha}(\vx,t) u_{\beta}(\vy,t)\right\rangle ,
\label{secondorder}
\end{eqnarray}
where the summation over repeated indices is implied. The explicit form of
the inertial operator ${\cal D}_{\alpha\,\beta}^{\gamma\,\delta}$, of scaling
dimension $-2+\xi$, can be derived using functional integrations by parts
\cite{Legacy} or the Dyson-Wyld equation \cite{RG1,ALM99}. It is not given
here for the sake of brevity. Under the assumption that a steady state sets
in, the inertial-range isotropic solution is sought in the time-independent
scaling form:
\begin{eqnarray}
C_{\alpha\,\beta}({\bf r})=C\,|\vr|^{\zeta_2}
\left[(d-1+\zeta_2)\delta_{\alpha\,\beta}-\zeta_2\,\hat{r}_\alpha\,\hat{r}_\beta
\right].
\label{Ansatz}
\end{eqnarray}
After the adoption of the Anstaz (\ref{Ansatz}), we define the integrals in
Eq.~(\ref{secondorder}) in the spirit dimensional regularization.
Taking the trace then gives the desired equation for $\zeta_2$:
\begin{eqnarray}
{\cal D}_{\alpha\,\alpha}^{\gamma\,\delta} \,C_{\gamma\,\delta}=
D\,C\,|\vr|^{-2+\xi+\zeta_2}\,
\frac{(d-1)(\zeta_2+\xi+d-2)}{(\zeta_2+\xi-2)}\, T(\xi,\zeta_2,\A) \, =0,
\label{realspace}
\end{eqnarray}
where
\begin{eqnarray}
T(\zeta_2,\xi,\A)& \equiv&
-(\A-1)^2(d+1)\xi\,\zeta_2+(\zeta_2+\xi-2)\left[d^2\,(\zeta_2+\A^2\,\xi)+
d\,\left(-\zeta_2+\zeta_2^2+\A^2\,\xi\,(\xi-1)\right)-(\zeta_2-\A\,\xi)
^2\right]-
\nonumber \\
&-& 4 (\A-1) (\A\eps-\A-d-1) \frac
{ \Gamma(1+\eps/2)\Gamma(1+d/2+\zeta_2/2)\Gamma(2-\eps/2-\zeta_2/2)
\Gamma(1+d/2)}
{\Gamma(d/2+\eps/2+\zeta_2/2) \Gamma(-\zeta_2/2) \Gamma(2+d/2-\eps/2)}.
\label{Gammas}
\end{eqnarray}
Equation (\ref{realspace}) has always a simple root for $\zeta_2=2-d-\xi$
and for all $\A\,\neq\,0\,,1$ a simple pole for $\zeta_2=2-\xi$.
It is worth stressing that the pole arises because of the nonlocal terms
in the inertial operator. The scaling $|\vr|^{2-\xi}$ is marginal and produces
logarithmic divergences at the edges of the inertial range which are reflected
in the presence of the pole.
In order to understand its physical meaning it is instructive to compare
Eq.~(\ref{realspace}) with the passive scalar case \cite{Falk1,GK}.
The inertial operator governing the pair correlation function again has
dimension $-2+\xi$; its action onto a power function yields
\begin{eqnarray}
{\cal D} |{\bf r}|^{-2+\xi+\zeta_2}=
D\,C\,|{\bf r}|^{-2+\xi+\zeta_2} (d-1)\,\zeta_2\,(\zeta_2+\xi+d-2).
\label{scalar}
\end{eqnarray}
The root $\zeta_2=2-d-\xi$ is again present. It corresponds to the
dimensional solution of the equation with the right-hand side is proportional
to $\delta({\bf r})$ (or a constant in momentum space). The
corresponding scaling behavior is admissible at scales much larger than
the integral one \cite{Falk1}. The second root $\zeta_2=0$,
a constant zero mode, is a consequence of energy conservation and as
such it is also a root of Eq.~(\ref{realspace}) for $\A=0$. Finally, setting
$\zeta_2=2-\xi$ in Eq.~(\ref{scalar}) solves the equation with a constant
right-hand side (or $\delta({\bf k})$ in the momentum space).
The pole for $\zeta_2=2-\xi$ in Eq.~(\ref{Gammas}) thus corresponds
to the scaling solution obtained by matching the forcing at large
scales. Its cancellation for $\A=0$ indicates the existence of a
constant flux solution in the presence of kinetic energy conservation.

The transcendental equation $T(\zeta_2,\xi,\A)=0$ has infinitely many
solutions. The leading admissible exponent behaves as $\zeta_{2}^{+}=
O(\xi)$ for small $\xi$. It can be obtained from Eq.~(\ref{Gammas})
within perturbation theories in $1/d$ or $\xi$:
\begin{eqnarray}
\zeta_{2}^{+} &=&
-\A^2\eps+ \frac {\A^2\eps}{d}
\left\{
\frac {(\A-1)\Gamma \left(1+{\eps}/2 \right)
\Gamma \left(1+(\A^2-1)\eps/2 \right)}
{\Gamma \left(1+{\A^2\eps}/2 \right)} -
\frac {\eps (\A+1) (\A^3\eps-\A^2\eps+ \A\eps+\A-\eps+1)} {(\A^2\eps-\eps+2)}
\right\}+O(1/d^{2})
\nonumber \\
&=& - \eps   \frac{\A^2(d-1)(d+2)}
{(d^{2}+\A^2+\A d-3)} -  \frac
{\eps^{2} \,\A^2 (d-1)} {2d(d^{2}+\A^2+\A d-3)^{2}}
\Bigl\{ d^{3} (\A+1)^{2} + (d^{2}-2d+4) (3\A^2+2\A+3) \Bigr\} +O(\eps^{3}).
\label{Expansions}
\end{eqnarray}
The nonlocality parameter $\A$ enters the perturbative expansions
from the first order. The fact reflects the balance between local and
nonlocal contributions in the zero-mode equation. It is worth noting
that up to first order in $1/d$ the exponent of the
linearized Navier-Stokes and of the magnetic, local, model coincide.

The next-to-leading correction exponent has the form
$\zeta_{2}^{-}=2-\eps+\A(\A-1)(d+2)\eps/(d^{2}+\A^2+\A d-3)+O(\eps^{2})$.

Nonperturbative solutions of Eq.~(\ref{realspace}) can be obtained
analytically  only in the case $\A=0$, where $\zeta_{2}^{+}= 0$
and $\zeta_{2}^{-}=2-\eps$ exactly, and in the local case $\A=1$ when
it reduces to a third-order algebraic equation. Besides the subleading
solution $\zeta_2^{-}=2-\xi$, the result of Ref. \cite{V96}
is recovered. In general, the behaviour of the exponents
$\zeta_{2}^{\pm}$ depends qualitatively on the value of $\A$.
Below we focus
our attention only on the behaviors of the leading exponent
$\zeta_{2}^{+}$  vs $\xi$ for $d=3$ and in the physically interesting
case $\A\in [-1,1]$. More detailed analysis will be reported elsewhere.

When $-0.581 <\A < 0.613$, the solution $\zeta_{2}^{+}$ exists and is
strictly negative within the entire interval $0 \le\eps\le 2$,
except for the case $\A=0$ when $\zeta_{2}^{+}$ vanishes identically.
For the other values of $\A$,
the real solution exists only in the subinterval $0 \le\eps\le \eps_{c}$,
($\eps_{c}(\A,d)\le 2$ is some critical value) where it is again
strictly  negative. For $\eps\ge\eps_{c}$,
the solution $\zeta_{2}^{+}$ coalesces with the closest unphysical branch,
$\zeta_{2}=-d+O(\eps)$, and they both become complex: the effect known for
the magnetic model \cite{V96}, where $\eps_{c}(1,3)=1$. It was argued in
Refs. \cite{V96,LM99} that this critical value of $\eps$ coincides with the
threshold for the dynamo effect (exponential growth of the pair correlation
function).

It is worth noticing that in the small subinterval, $-0.613< \A< -0.581$,
both solutions again become real near to $\eps=2$ (see Fig.~1).

Coefficients of the expansions in $\eps$ for $\zeta_{2}^{\pm}$ diverge at
$(d^{2}+\A^2+\A d-3) =0$ (this can happen only for $d\le2$, and for $d=2$
only for $\A=-1$). Nonperturbative analysis shows that in the region
$(d^{2}+\A^2+\A d-3) <0$, the effective diffusivity coefficient is negative
at large scales, which makes the system unstable with respect to any small
perturbation. In the limit $(d^{2}+\A^2+\A d-3) \to0$, the critical value
$\eps_{c}$ vanishes, and the solution for the leading inertial-range
exponent disappears. We thus conclude that the solution $\zeta_2=-\xi$,
reported in Ref. \cite{YK} for $d=2$ and $\A=-1$, in fact corresponds
to the large-scale exponent $\zeta_2=2-d-\xi$ mentioned above and not
to the inertial range.

Nonperturbative analysis can be repeated for the anisotropic sectors of
the pair correlation function; see Refs. \cite{LM99,ALM99,Arad} for $\A=1$
and \cite{LA,Arad2} for $\A=0$. Anisotropic contributions can be excited,
e.g., in the presence of a constant background field \cite{LM99,ALM99}.
The correlation can be expanded in components of definite total angular
momentum $j$ with radial dependence of the powerlike form
(parity considerations allow only even $j$ in this expansion):
   \begin{equation}
   {\cal C}_{\alpha\,\beta}({\bf r})\simeq \sum_{j=0}^{\infty}
   C_{\alpha\,\beta}(j)\,\,|\vr|^{\zeta_2(j)} \,\, P_{j}(\hat{\vr}).
   \label{scaling}
   \end{equation}
However, up to first order in $\xi$ these exponents can be most conveniently
calculated using the renormalization group and operator product expansion.
We refer the reader to the literature \cite{RG,RG1,ALM99,RG3} for details
and present only the result:
\begin{eqnarray}
\zeta_{2}^{(j)}= (j-2) +
\frac {\eps (d+2)} {(\A^2+d^2+\A d-3)(2j+d-6)(2j+d-4)(2j+d-2)}
\nonumber \\
\times \Bigl\{
(j^4 + 4j^3d + 2j^2d^2 - 10j^3 -22j^2d -6jd^2
+35j^2 +38jd +6d^2 -50j -24d+24)\A^2
\nonumber \\
+ (j-2) (j^2 +2jd +d^2 -5j -4d +4) [2\A(j+d-3)+(j-3)]
\Bigr\},
\label{corre3}
\end{eqnarray}
where $j\ge2$.
The dependence on the angular momentum indicates the existence of an
hierarchy of exponents growing with $j$ without an upper bound. The
result is in agreement with previous findings for the passive advection of
the scalar \cite{RG3} and magnetic \cite{LM99,ALM99,Arad} fields
and with the nonlocal case $\A=0$ \cite{LA,Arad2}.

The renormalization group and operator product analysis is also extended to
correlation functions of arbitrary order. The key role is played by the
dimensions $\Delta[n,j]$ associated with the $j\,$th rank tensor composite
operators $u_{\alpha_{1}}\dots u_{\alpha_{j}}\,({\bf u}^2)^{l}$, where
$n\equiv 2l+j$. Let ${\cal D}_{n,p}(\vx,\vy)$ be some equal-time scalar
correlation function of the $n\,$th and $p\,$th powers of the field $\vu$.
For the leading term in the $j\,$th anisotropic sector we obtain:
   \begin{eqnarray}
   {\cal D}_{n,p}(\vx,\vy)\propto (|\vr|/\ell)^{-\Delta_{n,j_n}-
   \Delta_{n,j_p}}\,(|\vr|/L)^{\Delta_{n,j}}\,P_j(\hat{\vr}),
   \label{zeta}
   \end{eqnarray}
where $j_k=0$ or 1 for even or odd $k$, respectively; cf. \cite{RG1,RG3}
for the scalar and \cite{ALM99} for the magnetic cases. Up to order
$O(\eps)$, we have obtained
\begin{equation}
\Delta[n,j]= \frac {n\eps}{2} +
\frac {\eps\, \A^2 [2n(n-1) - (d+1)(n-j)(d+n+j-2)]}
{2(d^{2}+\A^2+\A d-3)} +O(\eps^{2}).
\label{Dnp}
\end{equation}

This result, in particular, demonstrates the persistence of anisotropy
at small scales. Like in the scalar case \cite{Pumir}, dimensionless
ratios of the form $R_{2 n+1}\,\equiv\,{\cal D}_{2n+1,0}/{\cal D}_{2,0}^{(2n+1)/2}$
can be constructed using Eqs. (\ref{zeta}) and (\ref{Dnp}) and evaluated at
the dissipative scale $\ell$. In terms of the P\'eclet number
$\mbox{Pe}\propto (L/\ell)^{1/\xi}$ we obtain
\begin{eqnarray}
{\cal R}_{2k+1} \propto \mbox{Pe}^{\sigma_{2k+1}}, \quad
\sigma_{2k+1}= \frac{\A^2\, (d-1)\, (4k^{2}-d-2)}{2(d^{2}+\A^2+\A d-3)}
+O(\eps).
\label{ratio}
\end{eqnarray}
The ratio for $k=2$ (the so-called hyperskewness factor) diverges
as $\mbox{Pe}\to\infty$. This means that the conclusions drawn in the
scalar \cite{RG3} and magnetic \cite{ALM99,AHMM} cases regarding the
persistence of anisotropy at small scales remain valid in the presence of
the pressure term.

To summarize, the study of the pressure effects on the inertial-range scaling
behavior of a passive vector quantity advected by the rapid change model
reveals a double breakdown of the formal dimensional analysis. On one
hand, the leading scaling exponents are anomalous and governed by statistical
conservation laws (zero modes) of the inertial-range dynamics: a phenomenon
already known for a local vector (magnetic) model. On the other, as a
peculiar feature of nonlocality, the dimensionally regularized inertial
operator develops a pole corresponding to the scaling solution obtained
by formally matching the forcing at large scales. When the pressure term is
present, nonanomalous scaling of the pair structure function with the
dimensional exponent $2-\xi$ is realized only in the case when the
stretching term in Eq. (\ref{equation}) is absent and the energy
conservation enforces a constant flux solution.

The authors gratefully acknowledge valuable discussions with I.~Arad, A.~Celani,
A.~Kupiainen and M.~Vergassola.

The work of L.\,Ts.\,A., N.\,V.\,A. and A.\,V.\,R. was supported in part by
the RFFI Grant No.~99-02-16783 and GRACENAS Grant No.~E00-3-24.
A.\,M. was partially supported by the INFM PA project No.~GEPAIGG01.
P.\,M.\,G. was partially supported by EC Grant FMRX-CT98-0175.
P.\,M.\,G. thanks the ``Centro Internacional de Ciencias'' of Cuernavaca
and the organizers of the workshop ``Scaling and universality in strongly
nonlinear systems'' during which a part of this work has been performed.


\begin{figure}
\centerline{\psfig{file=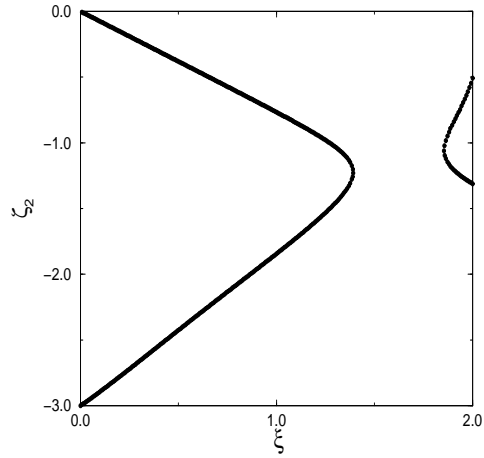,height=7cm,width=7cm}}
\caption{The leading exponent $\zeta_{2}^{+}$ (upper branches) and
the unphysical exponent $-d+O(\eps)$ (lower branches) {\it vs} $\xi$
for $d=3$ and $\A=-0.610$, within the small interval
$\A\in[-0.613,-0.581]$ (see text). The coalescence occurs at $\xi_{c}=1.390$
after which solutions become complex. The two branches becomes again
real for $\xi > 1.850$.}
\label{fig1}
\end{figure}



\end{document}